\documentclass[11pt]{article} 
\usepackage{fullpage,latexsym}

\begin{document}

\font\tenmsb=msbm8 scaled\magstep1\font\sevenmsb=msbm7 scaled\magstep1
\font\fivemsb=msbm5 scaled\magstep1\newfam\msbfam
\textfont\msbfam=\tenmsb\scriptfont\msbfam=\sevenmsb
\scriptscriptfont\msbfam=\fivemsb\def\Bbb#1{{\fam\msbfam\relax#1}}

\newcommand{\Bigotimes}{\displaystyle\bigotimes}
\newcommand{\Prod}{\displaystyle\prod}
\newcommand{\Sum}{\displaystyle\sum}
\newtheorem{theorem}{Theorem}
\newtheorem{lemma}{Lemma}	
\newenvironment{proof}{\noindent {\bf Proof }}{\hfill$\Box$\medskip}
\newtheorem{our_corollary}[theorem]{Corollary}

\title{A decision procedure for well-formed linear quantum cellular
automata\thanks{This research was supported by the ESPRIT Working Group 7097 RAND}}

\author{Christoph D\"urr,
	Huong L\^eThanh\\\small
	Universit\'e Paris-Sud, LRI, B\^at. 490\\\small
	 91405 Orsay Cedex, France\\\small
	 e-mail: \{durr,huong\}@lri.fr,
		http://www.lri.fr/$\sim$durr
\and
Miklos Santha\thanks{and by the French-Hungarian Research Program ``Balaton'' No.\ 94026 of the Minist\`ere des Affaires Etrang\`eres}\\\small
	CNRS, URA 410,
	Universit\'e Paris-Sud, LRI, B\^at. 490\\\small
	 91405 Orsay Cedex, France\\\small
	 e-mail: santha@lri.fr}

\date{}
\maketitle

\abstract{In this paper we introduce a new quantum computation model,
  the linear quantum cellular automaton.  Well-formedness is an
  essential property for any quantum computing device since it enables
  us to define the probability of a configuration in an observation as
  the squared magnitude of its amplitude.  We give an efficient
  algorithm which decides if a linear quantum cellular automaton is
  well-formed.  The complexity of the algorithm is $O(n^2)$ in the
  algebraic model of computation if the input automaton has continuous
  neighborhood.}

\textbf{key words:} quantum computation, cellular automata, de Bruijn graphs

\section{Introduction}

	In order to analyze the complexity of algorithms, computer
scientists usually choose some computational model, implement the
algorithm on it and count the number of steps as a function of the size
of the input. Different models, such as Turing machines (TM), random
access machines, circuits, or cellular automata can be used. They are
all universal in the sense that they can simulate each other with only
a polynomial overhead. However, these models are based on classical
physics, whereas physicists believe that the universe is better
described by quantum mechanics.

	Feynman \cite{Fey82,Fey86} and Benioff \cite{Ben82a,Ben82b}
were the first who pointed out that quantum physical systems are
apparently difficult to simulate on classical computers, suggesting
that there may be a gap between computational models based on
classical physics and models based on quantum mechanics.  Deutsch
\cite{Deu85} introduced the first formal model of quantum computation,
the quantum Turing machine (QTM).  He also described a universal
simulator for QTMs with an exponential overhead.  More recently,
Bernstein and Vazirani constructed a universal QTM with only a
polynomial simulation overhead \cite{BV97}.

	The power of QTMs was compared to that of classical
probabilistic TMs in a sequence of papers
\cite{Jozs91,DJ92,BB92,BV97}.  The most striking evidence that QTMs
can indeed be more powerful than probabilistic TMs was obtained by
Shor\cite{Sho94}, who built his work on an earlier result of Simon
\cite{Sim94}.  Shor has shown that the problems of computing the
discrete logarithm and factoring can be efficiently solved on a QTM,
whereas no polynomial time algorithm is known for these problems on a
probabilistic TM.

	Other quantum computational models were also studied. Yao
\cite{Yao93} has defined the quantum version of the Boolean circuit
model, and has shown that QTMs working in polynomial time can be
simulated by polynomial size quantum circuits.  Also, physicists were
interested in quantum cellular automata: Biafore \cite{Bia94}
considered the problem of synchronization, Margolus \cite{Mar94}
described space-periodic quantum cellular automata and Lloyd
\cite{Llo93,Llo94} discussed the possibility to realize a special type
of quantum linear cellular automaton (LQCA). However these models are
somehow different from the model of LQCA we consider in this article,
and the physical realizability of our model has not yet been studied.

	Well-formedness is an essential notion in quantum computation.
A quantum computational device is at any moment of its computation in
a superposition of configurations, where each configuration has an
associated complex amplitude.  If the device is observed in some
superposition of configurations then a configuration in the
superposition will be chosen at random.  The probability a
configuration will be chosen with is equal to the squared magnitude of
its amplitude.  Therefore it is essential that superpositions of unit
norm be transformed into superpositions of unit norm, or equivalently,
that the time evolution operator of the device preserve the norm. This
property is called the well-formedness.  In the case of a QTM,
Bernstein and Vazirani gave easily checkable local constraints on the
finite local transition function of the machine which were equivalent
to its well-formedness.  The existence of such relatively simple,
local criteria is due to the local nature of the evolution of a TM:
during a transition step only a fixed number of elements can be
changed in a configuration.

	In this paper we will define formally linear quantum cellular
automata and will give an efficient algorithm which decides if an LQCA
is well-formed. Our algorithm is of complexity $O(n^2)$ if the input
LQCA has continuous neighborhood (most papers in the literature in the
classical context deal only with such automata). The problem of
well-formedness in the case of an LQCA is much harder than in the case
of a QTM. One cannot hope for local conditions on the local transition
function as in the case of a QTM, since the transitions of a linear
cellular automaton are global: a priori no constant bound can be given
on the number of cells which are changing states in a step.  It turns
out that well-formedness is related to the reversibility of linear
classical cellular automata. Thus our work is closely related to the
decision procedure for reversible linear cellular automata of Sutner
\cite{Sut91}.

	In fact, quantum mechanics imposes an even stronger constraint
on any quantum computational device: its time evolution operator has
to be unitary.  For QTMs \cite{BV97}, space-periodic LQCAs
\cite{Dam96} and partitioned LQCAs \cite{Wat95} well-formedness
implies unitarity, but not for the model of LQCAs we consider
here. Building on the present algorithm we gave in a subsequent paper
\cite{DS96} an efficient procedure which decides if the evolution
operator of a LQCA is unitary.

	Watrous \cite{Wat95} has considered a subclass of LQCAs,
partitioned linear quantum cellular automata.  He has shown that a QTM
can be simulated by a machine from that class with constant slowdown,
and conversely, a partitioned LQCA can be simulated by a QTM with
linear slowdown. The efficient simulation of a general LQCA by a QTM
is left open in his paper.  As it is shown by Watrous, the problem of
well-formedness in the case of a partitioned LQCA is easy.  The local
transition function of a partitioned LQCA can be described by a finite
dimensional complex square matrix, and the automaton is well-formed if
and only if this finite matrix preserves the norm.  No analogous
result is known in the case of a general LQCA.

	Our paper is organized as follows. In section~2 we first
define linear cellular automata and give the basic notions of quantum
computation in a finite space.  Then we describe quantum linear
cellular automata, define the notion of well-formedness, and prove
that the inner product of two successor superpositions of
configurations can be reduced to the inner product of two finite
tensors.  In section~3 first we give an example which shows that the
trivial sufficient condition on the finite local transition function
is not necessary for well-formedness.  Then we describe the decision
procedure for well-formed quantum linear cellular automata, prove its
correctness, and analyze its complexity.  The procedure consists of
two separate algorithms, one which checks the unit norms, and another
which checks the orthogonality of the column vectors of the infinite
dimensional time evolution matrix of the automaton.  In section~4 we
describe a few open problems and finally in the appendix we give a
shorter proof of one of the main theorems of Watrous' paper.

\section{The computation model}
\label{sec-model}

\subsection{Linear cellular automata}

	A \emph{linear cellular automaton} (LCA) is a 4-tuple $A =
(\Sigma,q,N,\delta)$.  The \emph{cells} of the automaton are organized
in a line and are indexed by ${\Bbb Z}$.  $\Sigma$ is a finite
non-empty set of \mbox{\emph{(cell-)states}}.  At every step of the
computation, each cell is in a particular state.  The
\emph{neighborhood} $N=(a_1,\ldots,a_r)$ is a strictly increasing
sequence of signed integers for some $r \geq 1$, giving the addresses
of the neighbors relative to each cell. This means that the
\emph{neighbors} of cell $i$ are indexed by $i+a_1,\ldots,i+a_r$.  We
call $r = |N|$ the \emph{size} of the neighborhood.  Cells are
simultaneously changing their states at each time step according to
the states of their neighbors. This is described by the \emph{ local
transition function} $\delta : \Sigma^{|N|} \rightarrow \Sigma$.  If
at a given step the neighbors of a cell are respectively in states
$x_1, \ldots , x_r$ then at the next step the state of the cell will
be $\delta(x_1, \ldots, x_r)$.  The state $q \in \Sigma$ of $A$ is the
distinguished \emph{quiescent} state, which satisfies by definition
$\delta(q,\ldots,q)=q$.

	The set of \emph{configurations} is by definition
$\Sigma^{\Bbb Z}$, where for every configuration $c$, and for every
integer $i$, the state of the cell indexed by $i$ is $c_i$. The {\em
support} of a configuration $c$ is $ supp(c) = \{i \in {\Bbb Z} :
 c_i \neq q \}. $ A configuration $c$ will be called \emph{finite}
if it has a finite support.  We are dealing only with LCA's which work
on finite configurations.  Therefore from now on by 
\emph{configuration} we will mean \emph{finite configuration}.  The set
of configurations will be denoted ${\cal C}_A$.

	The local transition function induces a \emph{global
transition function}, $\Delta : {{\cal C}_A} \rightarrow {{\cal
C}_A}$, mapping a configuration to its \emph{successor}. For every
configuration $c$, and for every integer $i$, we have by definition
\[
		[\Delta(c)](i) = \delta(c_{i+N}),
\]
where $\delta(c_{i+N})$ is a short notation for
$\delta(c_{i+a_1},\ldots,c_{i+a_r})$. 

Configurations will often be represented by finite functions.
We call an \emph{interval} a finite subset of consecutive integers
$[j,k] = \{j, j+1, \ldots , k\}$ of ${\Bbb Z}$ for any $j$ and $k$ (if
$j > k $ this defines the empty interval $\emptyset$).  For our
purposes it will be convenient to deal with representations whose
domains are intervals.  Therefore for a configuration $c$, and for an
interval $I$, let $c_I$ be the restriction of $c$ to $I$. Also, let
$idom(c)$, the \emph{interval domain} of $c$, be the smallest interval
which contains $supp(c)$. For an interval $I = [j,k]$ with $j \leq k$,
we define $ext(I)$, the \emph{extension} of $I$ (with respect to the
neighborhood $N$) as the interval $[j - a_r, k - a_1]$. The extension
of $\emptyset$ is $\emptyset$. If $I = idom(c)$ then the support of
its successor $\Delta(c)$ is contained in $ext(I)$. Clearly, for every
configuration $c$ and intervals $I$ and $I'$, if $idom(c) \subseteq I$
and $ext(idom(c)) \subseteq I'$ then $c_I$ and $\Delta(c)_{I'}$
specify respectively $c$ and $\Delta(c)$.

We will call an LCA \emph{simple} if the elements of its neighborhood
form an interval, that is $a_r -a_1 = r-1$. In the literature LCA's
are often by definition simple.

	A LCA is \emph{trivial} if its neighborhood consist of a
single cell. We can suppose without loss of generality that this
single neighbor is the cell itself, that is $N=(0)$.

\subsection{Basic notions of quantum computation}

Let $E$ be a finite set and let us consider the complex vector space
${\Bbb C}^E$ with the usual inner product which is defined for vectors
$u,v \in {\Bbb C}^E$ by
\[
	\langle u, v \rangle = \sum_{e\in E} u(e) \cdot \overline{v(e)}.
\]
The vectors in ${\Bbb C}^E$ will be called \emph{superpositions} over
$E$, and for a superposition $u$ and an element $e \in E$, we will say
that $u(e)$ is the \emph{amplitude} of $e$ in that superposition.  The
norm $\| u \|$ of a superposition $u$ defined by this inner product is
\[
	\| u \| = \sqrt{ \sum_{e\in E} |u(e)|^2 } 
		= \sqrt{ \langle u, u\rangle}.
\] 
Two superpositions $u$ and $v$ are \emph{orthogonal}, in notation $u
\perp v$, if $\langle u, v \rangle = 0$. A superposition is
\emph{valid} if it has unit norm. If a valid superposition $u$ over
the set $E$ is \emph{observed} then one of the element of $E$ will be
chosen randomly and will be returned as the result of this
observation. The probability that the element $e$ is returned is
$|u(e)|^2$.  After the observation the superposition $u$ is changed
into the trivial superposition in which $e$ has amplitude 1 and all
the other elements 0.

Let $I$ be an interval, and for each $i \in I$, let $u_i$ be a
superposition over $E$. The \emph{tensor product} $\otimes_{i\in I}
u_i$ is a superposition over ${E^I}$, that is an element of the
complex vector space ${\Bbb C}^{E^I}$, where by definition, for all $x
\in E^I$,
\[
	\left[\Bigotimes_{i\in I} u_i\right] (x) = \prod_{i\in I} u_i(x_i).
\]
For our purposes the useful property of this operator is that the
inner product of two tensors is the product of the respective inner
products. Indeed, since $I$ is finite, we have
\begin{equation} 
	\left\langle  	\Bigotimes_{i\in I} u_i , 
			\Bigotimes_{i\in I} v_i \right\rangle =
        \prod_{i \in I} \langle u_i, v_i \rangle.
			\label{tensor-inner}
\end{equation}

\subsection{Linear quantum  cellular automata}

A linear quantum cellular automaton differs from a classical one in
the sense that the automaton evolves on a superposition of
configurations.  The local transition function $\delta$ maps the state
vector of a neighborhood into a superposition of new states, giving
the amplitude with which a cell moves into a specific state given the
state of its neighbors.

A \emph{linear quantum cellular automaton} (LQCA) is a 4-tuple $A =
(\Sigma,q,N,\delta)$, where the states set $\Sigma$ and the
neighborhood $N$ are as before.  It is called \emph{simple} if the
integers in $N$ form an interval.  The \emph{local transition
function} is $\delta : \Sigma^{|N|} \rightarrow {\Bbb C}^\Sigma$ such
that for every $(x_1, \ldots , x_r) \in \Sigma^r$, we have $\| \delta
(x_1, \ldots , x_r) \| > 0$. The distinguished \emph{quiescent} state
$q \in \Sigma$ satisfies for all $x\in \Sigma$
\[
	[\delta(q,\ldots,q)](x) = 
	\left\{
		\begin{array}{lll}
			1 & \textrm{if} & x = q,\\
			0 & \textrm{if} & x \neq q.\\
		\end{array}
	\right.
\]
Cells are again simultaneously changing their states at time steps but
the outcome of the changes is not unique.  If the neighbors of a cell
are respectively in states $x_1, \ldots , x_r$ then at the next step,
the cell will be in a superposition of states, where for every $y \in
\Sigma$, the state of the cell will be $p$ with amplitude
$[\delta(x_1, \ldots, x_r)](y)$.

The local transition function induces a global one, which maps a
superposition of configurations into its \emph{successor}
superposition. We call it the linear \emph{time evolution operator}
$U_A : {\cal C}_A \times {\cal C}_A \rightarrow \Bbb C$. For every $c,
d \in {\cal C}_A$, the automaton enters $d$ from $c$ in one step with
amplitude
\[
	U_A(d,c) = \prod_{i\in\Bbb Z} [\delta(c_{i+N})](d_i).
\]
This infinite product is well-defined since we deal with finite
configurations, so for all but a finite number of integers $i$,
$c_{i+N}=q^r$ and $d_i=q$. Therefore in the product only a finite
number of terms can be different from $1$. Moreover if there is an $i$
such that $c_{i+N}=q^r$ and $d_i\neq q$ then $U_A(d,c)=0$. Thus in
order to have non-zero transition amplitude it is necessary that
$idom(d)$ be contained in $ext(idom(c))$.

Let $I$ be any interval which contains $ext(idom(c))$. Then by the
previous observations and by definition of tensor product we have
\[
	U_A(d,c) = 
		\left\{
		\begin{array}{ll}
	\left[\Bigotimes_{i\in I} \delta(c_{i+N})\right](d_I) 
		& \textrm{if \ } idom(d) \subseteq I,\\
	0 & \textrm{otherwise.} 
		\end{array}
		\right.
\]
Clearly, superpositions of configurations form the Hilbert space
defined by
\[
	\ell_2({\cal C}_A) = \left\{ u \in {\Bbb C}^{{\cal C}_A} : 
		\sum_{c \in {\cal C}_A}
		u(c) \cdot \overline{u(c)} < \infty \right\},
\] 
with the inner product defined for $u_1,u_2 \in {\Bbb C}^{{\cal C}_A}$ by
\[
        \langle u_1, u_2 \rangle = 
		\sum_{c\in {\cal C}_A} u_1(c) \cdot \overline{u_2(c)}.
\]
As usual, $u_1$ and $u_2$ are \emph{orthogonal} (in notation $u_1 \perp
u_2)$ if $\langle u_1, u_2 \rangle = 0$.

As in the finite case, a superposition $v$ of configurations is
\emph{valid} if $\| v \| = \sqrt{ \langle v, v\rangle} = 1$. Also, as
in the finite case, if an LQCA is \emph{observed} in a valid
superposition of configurations $v$, the result of the observation
will be the configuration $c$ with probability $|v(c)|^2$. Immediately
after the observation whose outcome is $c$, the automaton will change
its superposition into the classical one which gives amplitude 1 to
$c$ and 0 to all the others.

We want to have valid superpositions of configurations at each moment
of the computation in order to associate the above probabilities to an
observation.  The initial configuration of the automaton is clearly
valid.  Therefore we say that the LQCA $A$ is \emph{well-formed} if its
time evolution operator $U_A$ preserves the norm.

It is not hard to see that $U_A$ preserves the norm if and only if its
column vectors are \emph{orthonormal}, that is they have \emph{unit
norms} and they are \emph{pairwise orthogonal}.  We will denote the
column vector of index $c$ by $U_A(\cdot, c)$.  In the next chapter we
will give an algorithm which decides if the column vectors of $U_A$
are orthonormal.  An important technical tool in the correctness of
the algorithm will be the generalization of equality
(\ref{tensor-inner}) to successor superpositions of configurations in
the infinite Hilbert space. This is stated in the following lemma.

\begin{lemma} \label{lem-inner}
Let $c$ and $c'$ be configurations and let $I$ be an interval such that
$ext(idom(c)) \cup ext(idom(d)) \subseteq I$. Then we have
\[
        \langle U_A(\cdot,c), U_A(\cdot,c') \rangle =
\prod_{i \in I} \langle \delta(c_{i+N}), \delta(c'_{i+N}) \rangle.
\]
\end{lemma}
\begin{proof}
\\[1em]
\hspace*{2em}$
        \langle U_A(\cdot,c), U_A(\cdot,c') \rangle
		=	$
\begin{eqnarray}
        &=&
         \Sum_{d \in {\cal C}_A} U_A(d,c) \cdot \overline{U_A(d,c')}
		\label{e1}\\
        &=& \Sum_{\scriptsize\begin{array}{c}
			d \in {\cal C}_A,\\ 
			supp(d) \subseteq I
		  \end{array}}
         \left[\Bigotimes_{i\in I} \delta(c_{i+N}) \right](d_I)
         \cdot \overline{
		\left[\Bigotimes_{i\in I} \delta(c'_{i+N})\right](d_I)}
		\label{e2}\\		
        &=& \Sum_{d' \in \Sigma^I}
        \left[\Bigotimes_{i\in I} \delta(c_{i+N}) \right](d')
         \cdot \overline{
		\left[\Bigotimes_{i\in I} \delta(c'_{i+N}) \right](d')}
		\label{e3}\\
        &=& \left\langle \Bigotimes_{i\in I} \delta(c_{i+N}),
         \Bigotimes_{i\in I} \delta(c'_{i+N}) \right\rangle
		\label{e4}\\
        &=& \Prod_{i \in I} \langle \delta(c_{i+N}), \delta(c'_{i+N}) \rangle.
		\label{e5}
\end{eqnarray}
The equations are justified in the following manner: (\ref{e1}) by
definition of the inner product, (\ref{e2}) by the choice of $I$,
(\ref{e3}) by identification of $d_I$ with $d'$, (\ref{e4}) by
definition of the tensor product and (\ref{e5}) by
equation~(\ref{tensor-inner}).
\end{proof} \medskip

We have the immediate corollary:
\begin{our_corollary} \label{cor-norm}
Let $c$ be a configuration and let $I$ be an interval such that\\
$ext(idom(c)) \subseteq I$. Then we have
\[
        \| U_A(\cdot,c) \| =
\prod_{i \in I} \| \delta(c_{i+N}) \|.
\]
\end{our_corollary}

\section{A decision procedure for well-formed LQCAs}
\label{sec-decision}

\subsection{Trivial LQCAs}

It is easy to give sufficient and necessary conditions for the
well-formedness of a trivial LQCA which are easily checkable on the
local transition function.  
\begin{lemma} 					\label{lem-suff-nec}
Let $A = (\Sigma,q,(0),\delta)$ be a trivial LQCA. Then $A$ is
well-formed if and only if for every $x,y \in \Sigma$ with $x\neq y$
\begin{equation}				
\label{cond-ortho}
		\delta(x) \perp \delta(y),
\end{equation}
and for every $x\in\Sigma$
\begin{equation}
\label{cond-norm}
		\| \delta(x) \| = 1.
\end{equation}
\end{lemma}
\begin{proof}
	For every $x\in\Sigma$ let $c^x$ be the configuration which is
$x$ at cell $0$ and quiescent elsewhere. Then for every $x,y \in
\Sigma$ we have $ \langle \delta ( x), \delta ( y) \rangle = \langle
U_A(\cdot, c^x), U_A(\cdot, c^y) \rangle $.  Thus if $A$ is well-formed
conditions~(\ref{cond-ortho}) and~(\ref{cond-norm}) hold.
 
	For the converse suppose that both conditions are satisfied.
Then corollary~\ref{cor-norm} implies that the columns of $U_A$ have
unit norm.  Now we show that for any two distinct configurations $c$
and $c'$, the associated columns of the evolution operator are
orthogonal. Since $c$ and $c'$ are different there exist a cell $i$,
such that $c_i \neq c'_i$. Thus $\delta(c_i) \perp \delta(c'_i)$ by
condition~(\ref{cond-ortho}) and $U_A(\cdot, c) \perp U_A(\cdot, c')$
by lemma~\ref{lem-inner}.
\end{proof} \medskip

For non-trivial LQCAs condition~(\ref{cond-ortho}) can never hold since
when $|N|>1$ we can not have $|\Sigma|^{|N|}$ independent vectors in a
space of dimension $|\Sigma|$. 

But condition~(\ref{cond-norm}) still implies that the column vectors
have unit norm by corollary~\ref{cor-norm}.  The following example
shows that this condition is not necessary.

Let $B = (\{q,p\},q,(0,1),\delta)$ be an LQCA with the local transition
function defined as follows. For $x \in \{q,p\}$, we define the
superposition $|x\rangle$ over $\{q,p\}$ by
\[
|x\rangle(y) =
\left\{
\begin{array}{lll}
1 & \textrm{if} & x = y,\\
0 & \textrm{if} & x \neq y.\\
\end{array}
\right.
\]
Then $\delta$ is defined as:

\[\begin{array}{ccccccc}
        \delta(q,q) &=& |q\rangle, &&
        \delta(q,p) &=& \frac1{2}|q\rangle, \\
        \delta(p,q) &=& 2|p\rangle, &&
        \delta(p,p) &=& |p\rangle.
  \end{array}
\]
In every configuration the number of pairs $qp$ is equal to the number
of pairs $pq$, therefore for all configurations $c,d$ we have
\[
	U_B(d,c) =
		\left\{
		\begin{array}{lll}
			1 & \textrm{if} & c = d,\\
			0 & \textrm{if} & c \neq d.\\
		\end{array}
		\right.
\]
Thus the time evolution matrix $U_B$ is just the identity, and $B$ is
well-formed.  However, $\delta(q,p)$ and $\delta(p,q)$ do not have
unit norm.

Nevertheless we can always transform a well-formed LQCA
$A=(Q,q,N,\delta)$ into an LQCA $A'=(Q,q,N,\delta')$ such that
$U_A=U_{A'}$ and $A'$ satisfies condition~(\ref{cond-norm}).  We
simply renormalize the local transition function for all
$w\in\Sigma^{|N|}$ by defining $\delta'(w) = \delta(w)/
\|\delta(w)\|$. Then for every configurations $c,d$ and interval $I$
containing $ext(idom(c))$ and $idom(d)$ we have
\begin{eqnarray*}
	U_{A'}(d,c) 	
&=& 	\left[ \Bigotimes_{i\in I} \delta'(c_{i+N}) \right](d_I)	\\
&=&	\prod_{i\in I} [\delta'(c_{i+N})] (d_i)			\\
&=&	\prod_{i\in I} \frac	{[\delta(c_{i+N})] (d_i)}
				{\|\delta(c_{i+N}) \|}		\\
&=&	\frac{	\prod_{i\in I}	[\delta(c_{i+N})] (d_i)}
	     {	\prod_{i\in I} \|\delta(c_{i+N}) \|}		\\
&=&	\frac{	\prod_{i\in I}	[\delta(c_{i+N})] (d_i)}
	     {			 \| U_A(\cdot,c) \|}		\\
&=&	\frac{	\prod_{i\in I}	[\delta(c_{i+N})] (d_i)}
	     			1				\\
&=& 	U_A(d,c).
\end{eqnarray*}

The following lemma establishes a particular property of trivial LQCAs
which is not true in general.
\begin{lemma}						\label{triv-uni}
Let $A = (\Sigma,q,(0),\delta)$ be a trivial LQCA. If $A$ is
well-formed then $U_A$ is unitary. 
\end{lemma}
\begin{proof}
Suppose $A$ is well-formed.  By the previous lemma $\delta$ is
described by a unitary matrix. Let $\delta^{-1}$ be the local function
described by the inverse of this matrix, that is for all
$x,y\in\Sigma$ we have $[\delta^{-1}(y)](x) =
\overline{[\delta(x)](y)}$. Let $A'$ be the trivial LQCA $(\Sigma, q,
(0), \delta^{-1})$. Clearly $U_{A'} U_A = U_A U_{A'} = I$, which
concludes the proof.
\end{proof} \medskip

\subsection{The algorithm}

Before giving the algorithm, let us discuss the size of the input,
that is the size of an LQCA $A = (\Sigma,q,N,\delta)$. It is clearly
dominated by the size of the description of $\delta$. We will work in
the algebraic computational model, where by definition complex numbers
take unit space, arithmetic operations and comparisons take unit time.
Then $\delta$ can be given by a table of size $|\Sigma|^{r+1}$, when
the neighborhood is of size $|N| = r$. Therefore we define the 
\emph{size} of the automaton $n = |\Sigma|^{r+1}$, and we will do the
complexity analysis of our algorithm as a function of $n$.

Our main theorem is an immediate consequence of
Theorems~\ref{thm-norm} and~\ref{thm-ortho}.

\begin{theorem} \label{thm-main}
  There exists an algorithm $P$ which takes a simple LQCA as input,
  and decides if it is well-formed. The complexity of the algorithm is
  $O(n^2 )$.
\end{theorem}

What can we say about the well formedness of an LQCA which is not
necessarily simple? Let $A = (\Sigma,q,N,\delta)$ be an LQCA of size
$n$ whose neighborhood is $N = (a_1,\ldots,a_r)$. We can transform $A$
into a simple LQCA $A' = (\Sigma,q,N',\delta')$ such that $A$ and $A'$
have the same time evolution operator.  This can be done by taking as
neighborhood $N' = (a_1, a_1 +1, a_1+2, \ldots , a_r)$, and making the
local transition function $\delta'$ independent from the new neighbors
in $N'$. Then we can run $P$ on $A'$.

The size of $A'$ will depend also on another parameter, on the
\emph{span} $s$ of $A$ which is defined as $s = a_r -a_1 +1$. Since
$|N'| = s$, the size of $A'$ will be $n' = |\Sigma^{s+1}| =
n^{(s+1)/(r+1)}$.  Let us define the \emph{expansion factor} $e$ of
$A$ as $e = {(s+1)/(r+1)}$. Then the time taken by $P$ will be $O(n'^2
) = O(n^{2e} )$. We have therefore the following corollary:

\begin{our_corollary} \label{cor-main}
There exists an algorithm which takes an LQCA with expansion factor $e$
as input, and decides if it is well-formed. The complexity of the
algorithm is $O(n^{2e} )$.
\end{our_corollary}

\subsection{Unit norms of column vectors}

In this chapter we will give an algorithm which decides if the column
vectors of the time evolution operator have unit norms. Let $A =
(\Sigma,q,N,\delta)$ be a simple LQCA whose neighborhood is of size
$r$. We define an edge weighted directed de Bruijn graph $G_A =
(V,E,w)$ with vertex set $V = \Sigma^{r-1}$, edge set $E = \{(xz,zy):
x,y\in\Sigma, z\in\Sigma^{r-2}\}$ and with weight function
$w:E\rightarrow {\Bbb R}$ defined by $w((xz,zy)) = \| \delta(xzy) \|$.
The unweighted version of this graph was defined by Sutner in
\cite{Sut91}.  A \emph{path} is a sequence $p = (v_0,\ldots,v_k)$ of
vertices such that for $0 \leq i \leq k-1$, we have
$(v_{i},v_{i+1})\in E$.  The \emph{weight} $w(p)$ of a the path $p$ is
\[
	\prod_{0 \leq i \leq k-1} w((v_{i},v_{i+1})).
\]
We call the path $(v_0,\ldots,v_k)$ a \emph{cycle} if $v_0=v_k$ and $k
> 0$. If in addition, $v_0 = q^{r-1}$ then it is called a
\emph{q-cycle}.  Our algorithm is based on the following lemma.

\begin{lemma} \label{lem-norm} 
The column vectors of $U_A$ have unit weight if and only if the weight
of all $q$-cycles in $G_A$ is 1.
\end{lemma}
\begin{proof}
Let $T$ denote the set of $q$-cycles of $G_A$.  We define a mapping $M
: {\cal C}_A \rightarrow T$.  Let $c$ be a configuration with interval
domain $I = [j,k]$.  Let $t = k-j $, and for $i = 0, 1, \ldots , k-j$,
let $x_i = c_{j+i}$.  Then by definition
\[
M(c) = (q^{r-1}, q^{r-2}x_0, q^{r-3}x_0x_1, \ \ldots \ ,
x_0x_1 \ldots x_{r-2} , \ \ldots \ , x_tq^{r-2}, q^{r-1}).
\]
We have then
\[
	\begin{array}{rcl@{\hspace*{2em}}r}
	\| U_A(\cdot,c) \| &=& 	
		\|\delta(q^{r-1}x_0)\| \cdot 
		\|\delta(q^{r-2}x_0x_1)\| \cdot
				\ldots \cdot
		\|\delta(x_tq^{r-1})\|	
				&\mbox{by corollary~\ref{cor-norm}}  \\[.2em]
&=& 	\multicolumn{2}{l}{
		\|\delta(q^r)\| 		\cdot
		\|\delta(q^{r-1}x_0)\| 		\cdot 
				\ldots 		\cdot
		\|\delta(x_tq^{r-1})\|		\cdot
		\|\delta(q^r)\| 
	}
				\\[.2em]
	&=& w(M(c)).		&\mbox{by definition}  
	\end{array}
\]
Since the mapping $M$ is clearly surjective
the statement of the lemma follows.
\end{proof} \medskip

Verifying if all column vectors of $U_A$ are of unit norm is now
reduced to checking if all $q$-cycles in $G_A$ are of unit weight.
The algorithm we give now will just do that.

\begin{theorem} \label{thm-norm}
  There exists an algorithm $R$ which takes a simple LQCA $A =
  (\Sigma,q,N,\delta)$ as input, and decides if the column vectors of
  the time evolution operator $U_A$ have all unit norm.  The
  complexity of the algorithm is $O(n^2)$.
\end{theorem}

\begin{proof}
	Algorithm $R$ will construct the graph $G_A$ of
lemma~\ref{lem-norm} and then determines if it has a $q$-cycle of
weight different from 1.  This will be done by two consecutive
algorithms $R_1$ and $R_2$, from which the first will check if there
is a column of norm less than 1, and the second will check if there is
a column of norm greater than 1.  They are both modifications of the
Bellman-Ford single source shortest paths algorithm \cite[see also
\cite{CLR90}]{Bel58,FF62} (BF for short), when $q^{r-1}$ is taken for
the source.
They are based on the fact that BF detects negative cycles going
through the source. (Actually for our purposes any shortest paths
algorithm can be used which uses sum and min as arithmetic operations,
and which detects negative cycles. Floyd's algorithm would be another
example).

	Algorithm $R_1$ replaces every sum operation in BF by a
product operation, and initializes the shortest path estimate for the
source to $1$ (the shortest path estimates for the other vertices are
initialized to $\infty$ as in BF), and then runs it on $G_A$. This way
it computes the shortest paths when the weight of a path is defined as
the product of the edge weights. To see this let ${G_A}'$ be the same
graph as $G_A$ except the edge weights are replaced by their
logarithm. Then the weight of a shortest path in ${G_A}'$ given by BF
will be the logarithm of the shortest path in $G_A$ given by $R_1$.
For the same reason, negative cycles in ${G_A}'$ through the source
will correspond to $q$-cycles in $G_A$ with weight less than 1 which
will therefore be detected by $R_1$.

	Algorithm $R_2$ replaces every min operation in $R_1$ by max
and the default initial shortest path estimate $\infty$ by 0, and then
runs it on $G_A$. This way it computes the shortest paths when the
weight of a path is defined as the product of the reciprocal of the
edge weights. If we define ${G_A}'$ with negative logarithm edge
weights then negative cycles in ${G_A}'$ will correspond to cycles in
$G_A$ with weight greater than 1 and will be detected by $R_2$.

	The complexity of BF is $O(|V|\cdot|E|)$. In the graph $G_A$ we
have $|V| = |\Sigma|^{r-1}$. Every vertex has $|\Sigma|$ outgoing
edges, therefore $|E| = |\Sigma|^{r}$. Thus the complexity of the
algorithm $R$ is $O(|\Sigma|^{2r-1}) = O(n^2)$.
\end{proof} \medskip

In~\cite{Hoy96} H{\o}yer gave a linear time algorithm to decide if the
column vectors have all unit norm, improving the complexity of our
result.

\subsection{Orthogonality of column vectors}
Now we will build an algorithm which decides if the column vectors of
the time evolution matrix are orthogonal.  Let again $A =
(\Sigma,q,N,\delta)$ be a simple LQCA whose neighborhood is of size
$r$. We define the graph $H_A = (V,E)$ with vertex set $V =
\Sigma^{r-1} \times \Sigma^{r-1}$ and edge set
\[
	\begin{array}{rll}
	E = \{ &( (x_1z_1 , x_2z_2) , (z_1y_1 , z_2y_2) ) : \\&
	x_1,x_2,y_1,y_2 \in\Sigma, z_1,z_2 \in\Sigma^{r-2}, \delta(x_1
	z_1 y_1) \not\perp \delta(x_2 z_2 y_2) &\}.
	\end{array}
\]
For a path $p = ((u_0,v_0),\ldots,(u_k,v_k))$ of $H_A$, let $p_1 =
(u_0,\ldots,u_k)$, and $p_2 = (v_0,\ldots,v_k)$. Clearly, $p_1$ and
$p_2$ are paths in $G_A$. A cycle is called here a $q$-\emph{cycle} if
its first vertex is $(q^{r-1},q^{r-1})$.

\begin{lemma} \label{lem-ortho}
  The column vectors of $U_A$ are orthogonal if and only if $p_1 =
  p_2$ for every $q$-cycle $p$ in $H_A$.
\end{lemma}
\begin{proof}
  Let $L = \{ (c,c') \in {\cal C}_A \times {\cal C}_A : U_A(\cdot, c)
  \not\perp U_A(\cdot, c') \}$, and let $T$ denote the set of
  $q$-cycles.  We will define a mapping $M : L \rightarrow T$. For
  $(c,c') \in L$, let $I = [j,k]$ be an interval such that
  $ext(idom(c)) \cup ext(idom(c')) \subseteq I$.  Let $t = k-j $, and
  for $i = 0, 1, \ldots , k-j$ we define $x_i = c_{j+i}$, and $y_i =
  c'_{j+i}$. Then by definition
\[
   M(c,c') = ( (q^{r-1},q^{r-1}), (q^{r-2}x_0,q^{r-2}y_0), \ \ldots \ ,
   (x_tq^{r-2},y_tq^{r-2}), (q^{r-1},q^{r-1}) ).
\]
  Since $U_A(\cdot,c) \not\perp U_A(\cdot,c') $, lemma~\ref{lem-inner}
  implies that $M(c,c')$ is indeed a $q$-cycle in $H_A$. Also, it is
  clear that $M$ is surjective.  Finally $c \neq c'$ if and only if
  $M(c,c')_1 \neq M(c,c')_2$ since both are equivalent to the existence
  of $i \in I$ such that $x_i \neq y_i$.
\end{proof} \medskip

We can now affirm:

\begin{theorem} \label{thm-ortho}
  There exists an algorithm $S$ which takes a simple LQCA $A =
  (\Sigma,q,N,\delta)$ as input, and decides if the column vectors of
  the time evolution operator $U_A$ are orthogonal.  The complexity of
  the algorithm is $O(n^2 )$.
\end{theorem}

\begin{proof}
The algorithm $S$ constructs the graph $H_A$ and computes the strongly
connected component of the node $(q^{r-1},q^{r-1})$. By
lemma~\ref{lem-ortho} there exists two distinct configurations such
that the corresponding column vectors in $U_A$ are not orthogonal if
and only if in this component there is a vertex $(u,v)$ with $u \neq
v$. This can be checked easily.

Finding the strongly connected components in a graph can be done in
time $O(|E| )$ for example with Tarjan's algorithm \cite{Tar75}.  In
$H_A$ the size of number of vertices is $|V| = |\Sigma|^{2(r-1)}$.
Since every vertex has outdegree $|\Sigma|^2$, the number of edges is
$|E| = |\Sigma|^{2r}$. Therefore the complexity of the algorithm $S$
is $O(|\Sigma|^{2r} ) = O(n^2 )$.
\end{proof} \medskip

\section{Conclusion}
\label{sec-conclusion}

	It would be interesting to generalize results concerning
reversibility of a linear classical CA for the well-formedness of an
LQCA. For example a necessary condition for reversibility is the
notion of \emph{balancedness} of the local transition function
\cite{AP72}, which means that every state has the same number of
preimages.  How does balancedness generalizes to the quantum model?

	It remains open, as stated also by Watrous, whether a QTM can
simulate an LQCA with reasonable slowdown.   

\appendix{Partitioned linear quantum cellular automata}
\label{app-plqca}

	This appendix treats a special kind of LQCA, the partitioned
LQCA, which was the main topic of Watrous' paper [Wat95].  Our
aim is to provide a new, shorter proof to one of his results, based on
our approach.

	A \emph{partitioned linear quantum cellular automaton} (PLQCA)
is a LQCA $A=(\Sigma, q, N, \delta)$, which satisfies the following
restrictions:
\begin{enumerate} \item
	The state-set $\Sigma$ is the Cartesian product $\Sigma_1
	\times \cdots \times \Sigma_r$ of some finite non-empty sets
	$\Sigma_i$, $i\in \{1, \ldots, r \}$.
\item
	The local transition function $\delta : \Sigma^r \rightarrow
	{\Bbb C}^\Sigma$ is the composition of two functions, the 
	\emph{classical part} $\delta_p : \Sigma^r \rightarrow \Sigma$ and
	the \emph{quantum} part $\delta_Q : \Sigma
	\rightarrow {\Bbb C}^\Sigma$. For all $x_{i,j} \in \Sigma_j$,
	$i,j\in \{1, \ldots, r \}$, $\delta_p$ is defined by
\[
	\delta_p((x_{1,1},\ldots, x_{1,r}),
		(x_{2,1},\ldots, x_{2,r}), \ldots,
		(x_{r,1},\ldots, x_{r,r}) ) =
	(x_{1,1},x_{2,2},\ldots, x_{r,r}).  
\] 
\end{enumerate} 

The function $\delta_p$ defines a LCA $A_p = (\Sigma,N,
\delta_p)$ whose global transition function $\Delta_p$ is a
permutation on configurations such that for all $c\in {\cal C}_A$ and
$i \in \Bbb Z$,
\[
	[\Delta_p(c)](i) = \delta_p(c_{i+N}).
\]
Moreover, the time evolution operator $U_{A_p}$ of $A_p$ is a
unitary matrix since for all $c, d\in {\cal C}_A$, we have
\[
	U_{A_p}(d,c) = \left\{ \begin{array}{ll}
		1 & \mbox{if } \Delta_p(c) = d, \\
		0 & \mbox{otherwise.}
				\end{array} \right.
\]

The \emph{local transition matrix} $Q$ is the complex valued matrix,
indexed by $\Sigma$, defined for all states $x,y \in \Sigma$ by
\[
	Q(y,x) = [\delta_Q(x)](y).
\]
In fact $Q$ completely determines the local transition function
$\delta$.  

The function $\delta_Q$ defines a trivial LQCA $A_Q =
(\Sigma,q,(0),\delta_Q)$, with the time evolution operator
$U_{A_Q}$. Clearly, ${\cal C}_A$ and $q$ are respectively the set of
configurations and the quiescent state also of $A_p$ and $A_Q$.  It
turns out that unitarity of the local transition matrix is equivalent
to the unitarity of the time evolution operator, as stated in the
following theorem.
\begin{theorem}[{[Wat 95, theorem 3.1 and corollary 3.1]}]
\label{thm-watrous}
	Let $A$ be a PLQCA, $U_A$ its time evolution operator and $Q$
	its local transition matrix. Then the following statements are
	equivalent.
\begin{enumerate}
\item
	$Q$ is unitary.
\item
	$A$ is well-formed.
\item
	$U_A$ is unitary.
\end{enumerate}
\end{theorem}
\begin{proof}
The local transition function of $A$ is the composition of two
separate local transition functions, thus its time evolution operator
is also the composition of time evolution operators of the associated
LQCAs, that is $U_A = U_{A_Q} U_{A_p}$.  Since $U_{A_p}$ is unitary we
have that $U_A$ preserves the norm (resp.\ is unitary) if and only if
$U_{A_Q}$ preserves the norm (resp.\ is unitary).

The theorem follows from lemmas~\ref{lem-suff-nec} and~\ref{triv-uni}.
\end{proof}

\section*{Acknowledgements}
We are thankful to St\'ephane Boucheron, Bruno Durand, Richard Jozsa and
John Watrous for several helpful conversations.


\end{document}